\documentclass {elsart}
\usepackage{graphicx}
\usepackage[small]{subfigure,epsfig}

\usepackage{amsmath} \usepackage{amssymb} \usepackage{cite}

\theoremstyle{plain} 

\numberwithin{equation}{section} \numberwithin{table}{section}
\numberwithin{theorem}{section}
\numberwithin{lemma}{section}
\numberwithin{theorem}{section}

\newcommand{\cn}
\begin{document}
\begin{frontmatter}

\title{Elliptic solutions for a family of fifth order nonlinear evolution equations}
\author{Nikolai A. Kudryashov\corauthref{cor}},
\corauth[cor]{Corresponding author.} \ead{nakudr@gmail.com}
\author{Dmitry I. Sinelshchikov}

\address{Department of Applied Mathematics, National Research Nuclear University
MEPHI, 31 Kashirskoe Shosse,
115409 Moscow, Russian Federation}

\begin{abstract}
The family of fifth order nonlinear evolution equations is studied. Some traveling wave elliptic solutions are found. The classification of these exact solutions is given.

\end{abstract}


\end{frontmatter}

\section{Introduction}

Recently in Refs. \cite{Kudryashov2010,Kudryashov2010a,Kudryashov2011b} a method was introduced for classification of meromorphic exact solutions of nonlinear ordinary differential equations. This method allows us to classify and construct meromorphic exact solutions for a wide class of autonomous nonlinear ordinary differential equations in the explicit form. The main idea of the approach from \cite{Kudryashov2010,Kudryashov2010a,Kudryashov2011b} is to compare the Laurent series corresponding to solutions of ordinary differential equations with the Laurent series for the general form of possible meromorphic exact solution. The main advantage of the method from works  \cite{Kudryashov2010,Kudryashov2010a,Kudryashov2011b} is that we can construct and classify more general forms of exact solutions in comparison to existing methods for finding exact solutions \cite{Kudr88,Kudr90,Parkes01,Parkes02,Fan2000,Polyanin,Kudryashov05,Biswas01,Vitanov,Kudr08a,Vernov02,Kudryashov_CNSNS_2010}. Using this approach in the work \cite{Kudryashov2010}, meromorphic exact solutions of second-order differential equation were classified.  The authors of \cite{Kudryashov2011b} presented the classification of elliptic solutions of a third-order differential equation obtained with the help of the above mention method. This method was used for constructing and classifying traveling wave solutions of the Kawahara equation \cite{Kudryashov2010a} and the generalized Bretherton equation \cite{KudrPLA11} as well.

In this work we study traveling wave solutions of the equation that takes the form
\begin{equation}
u_{t}+\alpha u^{m}\,u_{x}+\mu\,u_{xx}+\beta\,u_{xxx}+\nu u_{xxxx}+\delta u_{xxxxx}=0.
\label{Kaw_1}
\end{equation}

Eq. \eqref{Kaw_1} at $m=1$ was obtained in \cite{Kudryashov2011} for the description of nonlinear waves in a viscoelastic tube.  Nonlinear wave processes described by \eqref{Kaw_1} at $m=1$ were studied numerically in \cite{Kudryashov2011a}. The meromorphic solutions of Eq. \eqref{Kaw_1} at $\alpha=\gamma=0$ were found and classified in \cite{Kudryashov2010a}. Elliptic solutions of Eq.\eqref{Kaw_1} at $\mu=\beta=\nu=0$ in the cases $m=1$ and $m=3$ in terms of the Jacobi elliptic function were found in \cite{Parkes02}. In \cite{Parkes02} elliptic solutions of Eq.\eqref{Kaw_1} in the case of $\mu=\nu=0$ and $m=2$ were obtained as well. The simple periodic solutions of Eq. \eqref{Kaw_1} at any $m$ were found in work \cite{Kudryashov2009}. However elliptic solutions of Eq. \eqref{Kaw_1} at $\nu\neq0$ and $\mu\neq0$ were not considered previously.

Using the traveling wave $u(x,t)=y(z)$, $z=x-C_{0}\,t$ in \eqref{Kaw_1} and integrating the result we obtain
\begin{equation}
C_{1}-C_{0}\,y+\frac{\alpha}{m}\,y^{m+1}+\mu\,y_{z}+\beta\,y_{zz}+\nu\,y_{zzz}+\delta\,y_{zzzz}=0
\label{rKaw_1}
\end{equation}
Here $C_{1}$ is an integration constant.

The aim of this work is to construct and classify elliptic solutions of Eq.\eqref{rKaw_1} at $m=1,2,4$. To this aim we use the approach suggested in \cite{Kudryashov2010,Kudryashov2010a,Kudryashov2011b}.

This paper is organized as follows. In section 2 we give a brief description of the method from \cite{Kudryashov2010,Kudryashov2010a,Kudryashov2011b} and present elliptic solutions of Eq.\eqref{rKaw_1} at $m=1$. In sections 3 and 4 we construct elliptic solutions of Eq.\eqref{rKaw_1} in the cases of $m=2$ and $m=4$ respectively. In the last section we discuss our results.

\section{Meromorphic solutions of the equation studied at $m=1$.}

Let us briefly describe the approach from works \cite{Kudryashov2010,Kudryashov2010a,Kudryashov2011b} for finding elliptic solutions of nonlinear ordinary differential equations.

We assume that solutions of Eq.\eqref{rKaw_1} can be presented in the
form of Laurent series in a neighborhood of a movable pole $z=z_{0}$
\begin{equation}
y(z)^{(l)}=\sum\limits_{k=0}^{\infty}\,a_{k}^{(l)}\,(z-z_{0})^{k-p}, \quad p,\,l\in \mathbb{N}
\label{L_general}
\end{equation}

In the case of $m=1$ solutions of Eq.\eqref{rKaw_1} admit one expansion ($l=1$) in the form \eqref{L_general}. At $m=2$ and $m=4$ solutions of Eq.\eqref{rKaw_1}  admit two ($l=2$) and four ($l=4$) expansions in the form \eqref{L_general} respectively.

Eq.\eqref{rKaw_1} is autonomous and thus without loss of generality we can set $z_{0}=0$. Using the residue theorem for an elliptic function we obtain the necessary condition for existence of the elliptic solutions. This condition is the following: the sum of coefficients of $z^{-1}$ in series \eqref{L_general} is zero. From the necessary condition follows that while Eq.\eqref{rKaw_1} possesses one expansion of type \eqref{L_general} the coefficient at $z^{-1}$ has to be zero.
In the general case we can present this necessary condition in the form
\begin{equation}
\sum_{l}a_{p-1}^{(l)}=0, \quad l=1,2,3, \ldots
\label{necessary_condition}
\end{equation}

From here and below in the case of one expansion of type \eqref{L_general} we omit the index $l$ in coefficients of the series.

We use the method from \cite{Kudryashov2010,Kudryashov2010a,Kudryashov2011b} for constructing the elliptic solutions for Eq.\eqref{rKaw_1}. The algorithm of this method is the following:

\begin{enumerate}

\item Construct the formal Laurent expansion of type \eqref{L_general} for a solution of Eq.\eqref{rKaw_1};

\item Check the necessary condition \eqref{necessary_condition} for existence of the elliptic solutions;

\item Take the general form of the possible elliptic solution for Eq.\eqref{rKaw_1} and
find the Laurent expansion for the possible solution;

\item Compare the formal Laurent expansion of solution for Eq.\eqref{rKaw_1}
that was found in the first step with the formal Laurent expansion
of the possible elliptic solution that was found in the
third step;

\item Solve the system of the algebraic equations obtained in the
fourth step and find the parameters of Eq.\eqref{rKaw_1} and parameter
of the possible elliptic solution.
\end{enumerate}

We can see that the second step of the algorithm coincides with the first two steps of the Painlev\'{e} test. The Laurent expansion for possible elliptic solution can be find by using textbook \cite{Abramowitz} or by using symbolic computation software.

Let us construct elliptic solutions of Eq.\eqref{rKaw_1} at $m=1$. Without loss of  generality we assume that $\alpha=6$, $\delta=-1$ in Eq.\eqref{rKaw_1}. Thus from Eq.\eqref{rKaw_1} we have
\begin{equation}
C_{1}-C_{0}\,y+3 y^{2}+\mu\,y_{z}+\beta\,y_{zz}+\nu\,y_{zzz}-y_{zzzz}=0
\label{rKaw_n=1}
\end{equation}

Eq.\eqref{rKaw_n=1} admits one Laurent expansion ($l=1$) of type \eqref{L_general} in a neighborhood of moveable fourth order (p = 4) pole. The Fuchs indices corresponding to expansion in the Laurent series are the following
\begin{equation}
j_{1}=-1,\quad j_{2}=12, \quad j_{3,4}=\frac{1}{2}\left(11\pm i\sqrt{159}\right)
\end{equation}

We see that the Fuchs index $j_{2}$ has a positive integer value. Thus the expansion
for the solution of Eq.\eqref{rKaw_n=1} can exist if $a_{12}$ is an arbitrary constant.

The expansion for the solution of Eq.\eqref{rKaw_n=1} in the Laurent series is the following
\begin{equation}
\begin{gathered}
y=\frac{280}{z^{4}}+\frac{280\nu}{11z^{3}}-\left(\frac{21\nu^{2}}{121}+
\frac{7\beta}{3}\right)\frac{20}{13z^{2}}
+\left(\frac{59\nu^{3}}{17303}+\frac{20\beta\nu}{1287}+\frac{\mu}{9}
\right)\frac{140}{23z}+ \\+
\ldots+a_{12}z^{8}+\ldots
\label{n=1_L_series}
\end{gathered}
\end{equation}

Series \eqref{n=1_L_series} corresponding to solution of Eq.\eqref{rKaw_n=1} exists in the case
\begin{equation}
\begin{gathered}
\frac{784354\mu^{2}\beta^{2}\nu^{2}}{3671430867}+
\frac{137745805669\mu\,\beta^{2}\nu^{5
}}{549479397531636}+\frac{4490297861411\mu\,\nu^{9}}{297960291083728988}+
\vspace{0.1cm}\\+
\frac{353857280651\mu^{2}\nu^{6}}{4029515581898664}+\frac{2781278216285\nu^{10}\beta}{
297960291083728988}+\frac {206626426493\nu^{6}\beta^{3}}{
3296876385189816}+\vspace{0.1cm}\\
+\frac {1673496887869\nu^{8}\beta^{2}}{
44324671400885304}+\frac{310\beta^{5}\nu^{2}}{65033397}
+\frac {194170261339\mu\,\beta\,{\nu}^{7}}{1704795053880204}+\vspace{0.1cm}\\
+\frac{163837\mu^{3}\beta\,\nu}{1498986567}+\frac {32794140175\mu^{2}\beta\,
\nu^{4}}{99905345005752}+\frac{80582613855300\nu^{12}}{99146286858110820757}+
\vspace{0.1cm}\\
+\frac {\mu^{2}\left( 5313276\beta^{3}+7643363\mu^{2} \right)}{506657459646}+
\frac{\nu\beta^{4}\left( 10286502336
\alpha+25164119899\nu^{3} \right)}{674361078788826}
-\vspace{0.1cm}\\
-{\frac {\nu}{85961304}} \left( 4961\,\nu\,\beta+19844\,
\mu+10737\,\nu^{3} \right)  \left( C_{0}^{2}-12 C_{1} \right)+\vspace{0.1cm}\\
+\frac {5\mu\nu^{3}}{1348722157577652} \left( 42262166548\beta^{3}+40586709267\mu^{2} \right)
=0
\label{n=1_compatibility_condition}
\end{gathered}
\end{equation}


The last equality is the compatibility condition for existence of the Laurent series
\eqref{n=1_L_series}. Series \eqref{n=1_L_series} does not exist if relation \eqref{n=1_compatibility_condition} is not satisfied.

In accordance with the classification of meromorphic solutions of autonomous ordinary differential equations presented in \cite{Kudryashov2010,Kudryashov2010a} there is one type of possible elliptic solution of Eq.\eqref{rKaw_n=1}. This are elliptic solutions corresponding to Laurent series \eqref{n=1_L_series}.


Let us construct elliptic solutions of Eq.\eqref{rKaw_n=1}. Taking into account the necessary condition for elliptic solutions to exist \eqref{necessary_condition} from series \eqref{n=1_L_series} we obtain
\begin{equation}
\begin{gathered}
\mu=-\frac{(531\nu^{2}+2420\beta)\nu}{17303}
\label{n=1_NC}
\end{gathered}
\end{equation}
Using the approach from \cite{Kudryashov2010,Kudryashov2010a} we find that the general form of the possible elliptic solution of Eq.\eqref{rKaw_n=1} corresponding to series \eqref{n=1_L_series} has the form
\begin{equation}
\begin{gathered}
y=-\frac{20}{13}\left(\frac{21\nu^{2}}{121}+\frac{7\beta}{3}\right)\wp(z,g_{1},g_{2})-
\frac{140\nu}{11}\wp'(z,g_{1},g_{2})+\\+
280\,\wp^{2}(z,g_{1},g_{2})-\frac{70g_{2}}{3}+h_{0}
\label{n=1_elliptic_general}
\end{gathered}
\end{equation}

Comparing expansion \eqref{n=1_L_series} with the Laurent series for \eqref{n=1_elliptic_general} we obtain
\begin{equation}
\begin{gathered}
h_{0}=\frac{C_{0}}{6}-\frac{1}{169}\left({\frac {4845\nu^{4}}{29282}}+{\frac {7\beta^{2}}{9}}
+{\frac {643\beta\nu^{2}}{726}}\right) , \vspace{0.1cm}\\
g_{2}=-\frac {\left( 121\beta+48\nu^{2} \right)
 \left( 2057\beta+504\nu^{2} \right)}{207843636}, \vspace{0.1cm}\\
g_{3}=-\frac {\nu^{2} \left( 769365\,\nu^{4}+
6912246\,\beta \nu^{2}+19399325\,\beta^{2}\right)}{934108684080}-\frac{41\beta^{3}}{2372760}
\label{n=1_elliptic_parameters}
\end{gathered}
\end{equation}
Taking into account \eqref{n=1_NC} from condition \eqref{n=1_compatibility_condition} we have
\begin{equation}
\begin{gathered}
C_{1}=-\frac {3\nu^{2} }{171424512006748}\left( 17324377875\,\nu^{6}+187507377750\,\nu^{4}\beta+\right.\\
\left.+709528086443\,\beta^{2}\nu^{2}+1078179110844\,\beta^{3} \right)+\frac{C_{0}^{2}}{12}-\frac{1860\beta^{4}}{199927}
\label{n=1_C1}
\end{gathered}
\end{equation}

Also using \eqref{n=1_NC} and \eqref{n=1_C1} we obtain the value of $a_{12}$
\begin{equation}
\begin{gathered}
a_{12}=\frac {\beta^{5} \left( 51717941\beta+
174854187\nu^{2} \right)}{1272898705720561920}
+\frac {27\nu^{10} \left(
322959854223\nu^{2}+5492718054164\beta \right)}{38819181898104129541859840}+\vspace{0.2cm}\\
\frac {\nu^{4}\beta^{2} \left(
14947544754846\beta\nu^{2}+4344006788100\nu^{4}+
26834189262353\beta^{2} \right)}{167038348444816621893120}
\end{gathered}
\end{equation}

Substituting $h_{0}$, $g_{2}$ and $g_{3}$ \eqref{n=1_elliptic_parameters} into formulae \eqref{n=1_elliptic_general} we obtain the elliptic solution of Eq.\eqref{rKaw_n=1}.

Let us note that at $\beta=-\frac{45\nu^{2}}{176}$ and $\beta=-\frac{35\nu^{2}}{121}$ elliptic solution \eqref{n=1_elliptic_general} degenerates to simple periodic solutions.

\section{Meromorphic solutions of the  equation studied at m=2.}

Let us consider the traveling wave solutions of the equation at $m=2$. From Eq. \eqref{rKaw_1} we have
\begin{equation}
C_{1}-C_{0}\,y+ \frac{\alpha}{3}y^{3}+\mu\,y_{z}+\beta\,y_{zz}+\nu\,y_{zzz}+\delta\,y_{zzzz}=0
\label{rKaw_n=2_1}
\end{equation}

Without loss of generality we assume that $\delta=-1$ and $\alpha=360$ in Eq.\eqref{rKaw_n=2_1}.
Thus from Eq.\eqref{rKaw_n=2_1} we obtain
\begin{equation}
C_{1}-C_{0}\,y+120\,y^{3}+\mu\,y_{z}+\beta\,y_{zz}+\nu\,y_{zzz}-y_{zzzz}=0
\label{rKaw_n=2}
\end{equation}

Eq.\eqref{rKaw_n=2} admits two different Laurent expansions ($l=2$) in a neighborhood of a moveable second order (p = 2) pole. The Fuchs indices corresponding to expansions of the solution in the
Laurent series are the following
\begin{equation}
j_{1}=-1,\quad j_{2}=8, \quad j_{3,4}=\frac{1}{2}\left(7\pm i\sqrt{71}\right)
\end{equation}

We see that the Fuchs index $j_{2}$ has a positive integer value. So expansions
for the solution of Eq.\eqref{rKaw_n=2} can exist if $a_{8}$ is an arbitrary constant.

Expansions for solutions of Eq.\eqref{rKaw_n=2}  in a neighborhood of a moveable second order pole  are the following
\begin{equation}
y^{(1,2)}=\pm \frac{1}{z^{2}}\pm\frac \nu{14z} \mp \frac{23\nu^{2}+98\beta}{5880} \pm \frac{1}{4} \left( \frac {\mu}{45}+\frac{3\nu^{3}}{1715}+\frac {\beta\,\nu}{126} \right) z+\ldots+a_{8}^{(1,2)}z^{6}+\ldots
\label{L_series_n=2}
\end{equation}

Series \eqref{L_series_n=2} corresponding to solutions of Eq.\eqref{rKaw_n=2} exist in the case
\begin{equation}
\begin{gathered}
\pm{\frac {\nu^{2}{\beta}^{3}}{2100}}\,\pm{\frac {\beta\,{
\mu}^{2}}{900}}+\frac {3\nu^{2}C_{1}}{28}\pm\frac {2323\nu^{3}\mu\,\beta}{617400
}\mp\frac {2C_{0}\,\mu\,
\nu}{315}\mp{\frac {31 C_{0}\,\nu^{4}}{27440}}\pm\vspace{0.1cm}\\ \pm \frac {577\nu^{8}}{
9882516}\mp\frac{C_{0}\,\nu^{2}\beta}{630}\pm
\frac {1289\nu^{4}{\beta}^{2}}{1234800}\pm\frac {31\nu^{2}\mu^{2}}{8400}\pm\frac{257\nu^{6}\beta}{540225}\pm\vspace{0.1cm}\\ \pm\frac {4381\nu^{5}\mu}{4321800}\pm\frac {\mu\,{
\beta}^{2}\nu}{630}=0
\label{n=2_compatibility_condition}
\end{gathered}
\end{equation}

In accordance with the classification of meromorphic solutions of autonomous ordinary differential equations presented in \cite{Kudryashov2010,Kudryashov2010a} there are different types of possible elliptic solutions of Eq.\eqref{rKaw_n=2}. The first type is elliptic solutions corresponding to one of the Laurent series \eqref{L_series_n=2}. The second type is elliptic solutions corresponding to both of the Laurent series \eqref{L_series_n=2}.

The necessary condition for existence of elliptic solutions  \eqref{necessary_condition}  in the case of one of the Laurent series \eqref{L_series_n=2} gives us $\nu=0$. In the case of elliptic solutions corresponding to both of the Laurent series \eqref{L_series_n=2} the necessary condition \eqref{necessary_condition} is automatically satisfied.


Let us construct elliptic solutions of Eq.\eqref{rKaw_n=2} corresponding to both of the series \eqref{L_series_n=2}. In this case we force that the compatibility conditions \eqref{n=2_compatibility_condition} are satisfied simultaneously. We see that this is correct at $C_{1}=0$ or at $\nu=0$.


In accordance with the method presented in \cite{Kudryashov2010,Kudryashov2010a} the possible elliptic solution of Eq. \eqref{rKaw_n=2} corresponding to both of the series \eqref{L_series_n=2} has the form
\begin{equation}
\begin{gathered}
y=-\left[\frac{\wp^{'}(z,g_{1},g_{2})+B}{\wp(z,g_{1},g_{2})-A}\right]^{2}-
\frac{\nu}{28}\frac{\wp^{'}(z,g_{1},g_{2})+B}{\wp(z,g_{1},g_{2})-A}
+2\wp(z,g_{1},g_{2})+h_{0}
\label{n=2_gen_elliptic}
\end{gathered}
\end{equation}
Here $A$, $B$ and $h_{0}$ are parameters that will be found later. Also we denote $\frac{d}{dz}$ by $'$.

First we consider the case of $C_{1}=0$.
Comparing expansion \eqref{L_series_n=2} with the Laurent series for \eqref{n=2_gen_elliptic} we obtain
\begin{equation}
\begin{gathered}
h_{0}=A=\frac{\nu^{3}-196\mu}{1680\nu},\quad B=0, \quad g_{2}=\frac {38416\,\mu^{2}+\nu^{6}-532\,\nu^{3}\mu}{141120\nu^{2}},\vspace{0.2cm}\\
g_{3}=\frac { \left(196\,\mu-\nu^{3} \right)  \left( 38416\,\mu^{2}-567\,\nu^{3}\mu+\nu^{6}
\right) }{296352000\nu^{3}},\quad
\beta=\frac{39\nu^{3}-1372\mu}{196\nu}
\label{n=2_elliptic_parameters}
\end{gathered}
\end{equation}

From the compatibility conditions \eqref{n=2_compatibility_condition} we have
\begin{equation}
C_{0}=\frac {134456\,\mu^{2}-4067\,\nu^{3}
\alpha+11\,\nu^{6}}{6860\nu^{2}}
\end{equation}

Solving the algebraic system of equations for parameters $h_{0}$, $A$, $B$, $g_{2}$, $g_{3}$ we obtain values for constant $a_{8}$ in the form
\begin{equation}
\begin{gathered}
a_{8}^{(1,2)}=\frac {1}{23897825280000\nu^{4}}\Big(\pm 1475789056\,\mu^{4} \mp 8605184
\,\nu^{3}\mu^{3}\mp \\ \mp 263424\,\nu^{6}\mu^{2} \mp 224\,
\nu^{9}\mu \pm \nu^{12}\Big)
\label{n=2_compatibility_condition1}
\end{gathered}
\end{equation}

Expressions \eqref{n=2_compatibility_condition1} are necessary conditions for existence of the Laurent series \eqref{L_series_n=2}. These conditions show us that the elliptic solution of Eq. \eqref{rKaw_n=2} contains only one arbitrary constant. In this case we can add the arbitrary constant $z_{0}$ to the variable $z$.

Taking into account formulae \eqref{n=2_elliptic_parameters} we have an elliptic solution of Eq.\eqref{rKaw_n=2} in the form
\begin{equation}
\begin{gathered}
y=-\left[\frac{\wp^{'}(z,g_{1},g_{2})}{\wp(z,g_{1},g_{2})-\frac{\nu^{3}-196\mu}{1680\nu}}\right]^{2}-
\frac{\nu\wp^{'}(z,g_{1},g_{2})}{28(\wp(z,g_{1},g_{2})-\frac{\nu^{3}-196\mu}{1680\nu})}
+2\wp(z,g_{1},g_{2})+h_{0}
\label{n=2_gen_elliptic1}
\end{gathered}
\end{equation}
where $g_{2}$, $g_{3}$ are defined by \eqref{n=2_elliptic_parameters}.

The elliptic solution \eqref{n=2_gen_elliptic} degenerates to simple periodic solutions at the following values of $\mu$
\begin{equation}
\mu=\frac{\nu^{3}}{476},\quad \mu=\frac{17\nu^{3}}{1372}, \quad \mu=\frac{\nu^{3}}{686}, \quad \mu=\frac{\nu^{3}}{56}
\end{equation}
In the case of $\mu=\frac{\nu^{3}}{476}$ we obtain the following simple periodic solution from solution \eqref{n=2_gen_elliptic}
\begin{equation}
\begin{gathered}
y=\frac{\nu^{2}\left(7-14\cos^{2}\left\{\frac{\sqrt{119}\nu z}{476}\right\}+\sqrt{119}\sin\left\{\frac{\sqrt{119}\nu z}{238}\right\}\right)}{13328\cos^{2}\left\{\frac{\sqrt{119}\nu z}{476}\right\}\sin^{2}\left\{\frac{\sqrt{119}\nu z}{476}\right\}}
\end{gathered}
\end{equation}
In this case parameters $A,h_{0},\beta,C_{0}$ can be written as
\begin{equation}
\begin{gathered}
h_{0}=A=\frac{\nu^{2}}{2856},\quad \beta=-\frac{178\nu^{2}}{833}, \quad C_{0}=\frac{705\nu^{4}}{1586032}
\end{gathered}
\end{equation}

At $\mu=\frac{17\nu^{3}}{1372}$ solution \eqref{n=2_gen_elliptic} degenerates to
\begin{equation}
\begin{gathered}
y=\frac{\nu^{2}}{49}\frac{e^{\frac{3\nu z}{14}}}{(e^{\frac{\nu z}{7}}-1)^{2}}
\end{gathered}
\end{equation}
Parameters $A,h_{0},\beta,C_{0}$ are defined by the following relations
\begin{equation}
\begin{gathered}
h_{0}=A=-\frac{\nu^{2}}{1176},\quad \beta=-\frac{2\nu^{2}}{7}, \quad C_{0}=-\frac{15\nu^{4}}{5488}
\end{gathered}
\end{equation}

In the case of $\mu=\frac{\nu^{3}}{686}$  from \eqref{n=2_gen_elliptic} we have
\begin{equation}
\begin{gathered}
y=\frac{\nu^{2}(e^{\frac{\nu z}{7}}+2e^{\frac{\nu z}{14}}-1)}{392(e^{\frac{\nu z}{14}}-1)^{2}}
\end{gathered}
\end{equation}
In this case parameters $A,h_{0},\beta,C_{0}$ can be written as
\begin{equation}
\begin{gathered}
h_{0}=A=\frac{\nu^{2}}{2352},\quad \beta=-\frac{41\nu^{2}}{196}, \quad C_{0}=\frac{15\nu^{4}}{19208}
\end{gathered}
\end{equation}

At $\mu=\frac{\nu^{3}}{56}$ solution \eqref{n=2_gen_elliptic} degenerates to
\begin{equation}
\begin{gathered}
y=\frac{\nu^{2}\left(7-\sin\left\{\frac{\sqrt{14}\nu z}{28}\right\}\right)}{1568 \cos^{2}\left\{\frac{\sqrt{14}\nu z}{56}\right\}}
\end{gathered}
\end{equation}
Parameters $A,h_{0},\beta,C_{0}$ are defined by the following relations
\begin{equation}
\begin{gathered}
h_{0}=A=-\frac{\nu^{2}}{672},\quad \beta=-\frac{127\nu^{2}}{392}, \quad C_{0}=-\frac{15\nu^{4}}{5488}
\end{gathered}
\end{equation}

In the case of $\nu=0$, the elliptic solution of type \eqref{n=2_gen_elliptic} degenerates to a simple periodic as well.

Let us construct elliptic solutions of equation \eqref{rKaw_n=2} that correspond to one of the series \eqref{L_series_n=2}. We see from \eqref{L_series_n=2}  that the necessary condition for existence of an elliptic solution is $\nu=0$. In this case the possible elliptic solution of Eq. \eqref{rKaw_n=2} corresponding to the first series from \eqref{L_series_n=2} is the following
\begin{equation}
\begin{gathered}
y=\wp(z,g_{2},g_{3})+h_{0}
\label{n=2_el_solution}
\end{gathered}
\end{equation}
Comparing the Laurent expansion for \eqref{n=2_el_solution} with the first expansion \eqref{L_series_n=2} at $\nu=0$ we have

\begin{equation}
\begin{gathered}
h_{0}=-\frac{\beta}{60},\quad \mu=0,\quad g_{2}=\frac{10C_{0}-\beta^{2}}{180}, \quad g_{3}=\frac{5\,C_{0}\beta-450C_{1}-\beta^{3}}{5400}
\label{n=2_el_solution_parameters}
\end{gathered}
\end{equation}

The coefficient $a_{8}$ is defined by the following relation
\begin{equation}
\begin{gathered}
a_{8}=\frac{(10C_{0}-\beta^{2})^{2}}{38880000}
\end{gathered}
\end{equation}

Substituting $g_{2},g_{3}$ and $h_{0}$ from \eqref{n=2_el_solution_parameters} into \eqref{n=2_el_solution} we obtain the elliptic solution of Eq.\eqref{rKaw_n=2} at $\mu=\nu=0$
\begin{equation}
\begin{gathered}
y=\wp\left(z,\frac{10C_{0}-\beta^{2}}{180},\frac{5\,C_{0}\beta-450C_{1}-\beta^{3}}{5400}\right)-
\frac{\beta}{60}
\label{n=2_el_solution1}
\end{gathered}
\end{equation}
This solution was previously obtained in work \cite{Kudryashov2010a}.

Thus we see that equation \eqref{rKaw_n=2} has elliptic solutions corresponding to one of the series \eqref{L_series_n=2} only at $\nu=\mu=0$. Elliptic solutions of Eq.\eqref{rKaw_n=2} at $\nu=\mu=\beta=0$ can be obtained form the formulae \eqref{n=2_el_solution1}.

Note that Eq.\eqref{rKaw_n=2} possesses the symmetry
\begin{equation}
y(z,-C_{1})=-y(z,C_{1})
\end{equation}
Thus the elliptic solution corresponding to the second series from \eqref{L_series_n=2} can be obtained from  solution \eqref{n=2_el_solution1}.

\section{Meromorphic solutions of the equation studied at m=4.}

Let us construct elliptic solutions of Eq.\eqref{rKaw_1} at $m=4$. In this case from Eq.\eqref{Kaw_1} we obtain
\begin{equation}
C_{1}-C_{0}\,y+ \frac{\alpha}{5}y^{5}+\mu\,y_{z}+\beta\,y_{zz}+\nu\,y_{zzz}+\delta\,y_{zzzz}=0
\label{rKaw_n=4_1}
\end{equation}

Without loss of generality we assume that $\delta=-1$ and $\alpha=120$ in Eq.\eqref{rKaw_n=4_1}.
Thus from Eq.\eqref{rKaw_n=4_1} we have
\begin{equation}
C_{1}-C_{0}\,y+24\,y^{5}+\mu\,y_{z}+\beta\,y_{zz}+\nu\,y_{zzz}-y_{zzzz}=0
\label{rKaw_n=4}
\end{equation}

Eq. \eqref{rKaw_n=4} possesses four different expansions in the Laurent series in a neighborhood of the first order moveable pole. The Fuchs indices corresponding to these expansions are the following
\begin{equation}
j_{1}=-1, \quad j_{2}=6, \quad j_{3,4}=\frac{1}{2}(5 \pm i \sqrt{39})
\end{equation}
We see that the coefficient $a_{6}$ in the Laurent series has to be arbitrary. Otherwise the Laurent expansions do not exist.

Expansions of solution of Eq.\eqref{rKaw_n=4} in a neighborhood of the moveable pole are the following
\begin{equation}
\begin{gathered}
y^{(1,2)}=\pm\frac{1}{z}\pm\frac{\nu}{20} \mp \frac{1}{600}\left(3\nu^{2}+10\beta\right)\,z+\ldots+a_{6}^{(1,2)}\,z^{5}+\ldots
\label{n=4_L_1}
\end{gathered}
\end{equation}
\begin{equation}
\begin{gathered}
y^{(3,4)}=\pm\frac{i}{z}\pm\frac{i\nu}{20} \mp \frac{i}{600}\left(3\nu^{2}+10\beta\right)\,z+\ldots+a_{6}^{(1,2)}\,z^{5}+\ldots
\label{n=4_L_2}
\end{gathered}
\end{equation}

The compatibility conditions for series \eqref{n=4_L_1} to exist are the following
\begin{equation}
\begin{gathered}
\pm \frac {189\nu^{6}}{100000}\pm \frac {\nu\beta
 \left( 10\nu\beta+35\mu+9\,\nu^{3} \right)}{750} \pm
 \frac {\mu\left( 99\nu^{3}+100\,\mu \right)}{3000} -\vspace{0.1cm}\\-
 \frac{\nu\left( 5 C_{1} \pm C_{0}\nu \right)}{10}=0
\label{n=4_CC}
\end{gathered}
\end{equation}

In the case of series \eqref{n=4_L_2} the compatibility conditions have the form
\begin{equation}
\begin{gathered}
\pm \frac {189i\nu^{6}}{100000}\pm \frac {i \nu\beta
 \left( 10\nu\beta+35\mu+9\,\nu^{3} \right)}{750} \pm
 \frac {i\mu \left( 99\nu^{3}+100\,\mu \right)}{3000} -\vspace{0.1cm}\\-
 \frac{i\nu\left( -5 i C_{1} \pm C_{0}\nu \right)}{10}=0
\label{n=4_CC_1}
\end{gathered}
\end{equation}

Using the necessary condition \eqref{necessary_condition} for elliptic solutions to exist we see that Eq.\eqref{rKaw_n=4} may admit three types of elliptic solutions. There are elliptic solutions corresponding to both of the series \eqref{n=4_L_1} or to both of the series \eqref{n=4_L_2}. Also Eq.\eqref{rKaw_n=4} may admit elliptic solutions corresponding to series \eqref{n=4_L_1} together with series \eqref{n=4_L_2}.



Let us construct elliptic solutions of Eq.\eqref{rKaw_n=4} corresponding to expansions \eqref{n=4_L_1}. The general form of possible elliptic solution of Eq.\eqref{rKaw_n=4} is the following \cite{Kudryashov2010,Kudryashov2010a}
\begin{equation}
\begin{gathered}
w=-\frac{\wp^{'}(z,g_{2},g_{3})+B}{2(\wp(z,g_{2},g_{3})-A)}+h_{0}
\label{n=4_E_general}
\end{gathered}
\end{equation}

Compatibility conditions \eqref{n=4_CC} are the same at $C_{1}=0$ or at $\nu=0$.
In the case of $C_{1}=0$, the algebraic system of equations for parameters $\mu,\beta,\nu,C_{0}$ is inconsistent. Thus we have to consider the case of $\nu=0$. Comparing the Laurent expansion of \eqref{n=4_E_general} with series \eqref{n=4_L_1} we find
\begin{equation}
\begin{gathered}
\mu=h_{0}=B=C_{1}=0, \quad A=-\frac{\beta}{60}, \quad g_{2}=\frac{\beta^{2}-10C_{0}}{120},\\
g_{3}=\frac{\beta(13\beta^{2}-150 C_{0})}{108000}
\label{n=4_El_coeffs}
\end{gathered}
\end{equation}
Coefficients $a_{6}^{(1,2)}$ are defined by the relations
\begin{equation}
\begin{gathered}
a_{6}^{(1,2)}=\mp\frac{\beta(18 C_{0}-\beta^{2})}{302400}
\end{gathered}
\end{equation}

Substituting $A,B,h_{0}$ from \eqref{n=4_El_coeffs} into formulae \eqref{n=4_E_general} we obtain an elliptic solution of Eq.\eqref{rKaw_n=4} at $\mu=\nu=C_{1}=0$
\begin{equation}
\begin{gathered}
w=-\frac{\wp^{'}(z,g_{2},g_{3})}{2(\wp(z,g_{2},g_{3})+\frac{\beta}{60})}
\label{n=4_E_general_1}
\end{gathered}
\end{equation}
where $g_{2},g_{3}$ are defined by \eqref{n=4_El_coeffs}. Solution \eqref{n=4_E_general_1} was obtained in \cite{Kudryashov2010a}.

In the case of series \eqref{n=4_L_2}, the compatibility conditions \eqref{n=4_CC_1} are the same at $C_{1}=0$ or at $\nu=0$ as well. In the case of $C_{1}=0$ algebraic system of equations for parameters $\mu,\beta,\nu,C_{0}$ is inconsistent again. For obtaining elliptic solutions corresponding to series \eqref{n=4_L_2} we take into account the following symmetries of Eq.\eqref{rKaw_n=4}
\begin{equation}
y(z,-C_{1})=-y(z,C_{1}), \quad y(z,iC_{1})=-iy(z,C_{1})
\label{n=4_symm}
\end{equation}

Using \eqref{n=4_El_coeffs} and \eqref{n=4_E_general_1} we have
\begin{equation}
\begin{gathered}
w=-\frac{i\wp^{'}(z,g_{2},g_{3})}{2(\wp(z,g_{2},g_{3})+\frac{\beta}{60})}
\label{n=4_E_general_11}
\end{gathered}
\end{equation}
where $g_{2},g_{3}$ are defined by \eqref{n=4_El_coeffs} again. Coefficients $a_{6}^{(3,4)}$ are given by the following relations
\begin{equation}
\begin{gathered}
a_{6}^{(3,4)}=\mp\frac{i\beta(18 C_{0}-\beta^{2})}{302400}
\end{gathered}
\end{equation}

In the case of elliptic solutions with poles of four different type, the algebraic system for the parameters is inconsistent.

We can consider the case of $\mu=\beta=\nu=0$. However while this condition is satisfied elliptic solutions of type \eqref{n=4_E_general} degenerate to simple periodic solutions.

\section{Conclusion}
In this paper we have studied elliptic traveling wave solutions for the family of fifth order nonlinear evolution equations. We have presented a classification of elliptic solutions of this family in the cases of $m=1$, $m=2$ and $m=4$. The explicit form of the elliptic solutions is given as well. In the case of $m=1$ we have obtained a new elliptic solution with one pole in the  parallelogram of periods. At $m=2$  a new elliptic solution with two poles in the parallelogram of periods is presented. We have shown that elliptic solution with one pole in the parallelogram of periods exists only in the case $\nu=\mu=0$. In the case of $m=4$ we have given elliptic solutions with two poles in the  parallelogram of periods. As well, for elliptic solutions with four poles in the parallelogram of periods the corresponding algebraic system is inconsistent.

\end{document}